# Towards sub-30nm Contacted Gate Pitch, Forked Contact and Dynamically-Doped Nanosheets to Enhance Si and 2D Materials Device Scaling

Aryan Afzalian, Zubair Ahmed and Julien Ryckaert
imec, Leuven, Belgium, aryan.afzalian@imec.be

**Abstract**

We propose a novel Forked-Contacts, Dynamically-Doped Multigate transistor as ultimate scaling booster for both Si and 2D materials in aggressively-scaled nanosheet devices. Using accurate dissipative DFT-NEGF atomistic-simulation fundamentals and cell layout extrinsics, we demonstrate superior and optimal device characteristics and invertor energy - delays down to sub-30-nm pitches, i.e., a 10 nm scaling boost compared to the nanosheet MOSFET references. **Keywords:** CMOS, CGP scaling, Si, TMD, 2d materials, dynamic-doping

## Introduction

The Dynamically-Doped (D2) Field-Effect Transistor is a novel device architecture that scales better than its MOSFET nanosheet (NS) counterpart [1], owing to the suppression of ungated extensions (spacers) from the device Contacted Gate-pitch (CGP) equation [1,2] and Fig. 1a. What used to be the NS chemically doped extensions are now electrically and dynamically-doped by the gate, i.e., a part of the channel. Hence, for a given CGP, the channel length $L$ in the D2FET is twice the spacer length ($L_{SPACER}$) longer than $L$ of a standard MOSFET, as it benefits from the full distance between the source (S) and drain (D) contact pads. The gate length $L_G$ value could even be larger than $L$, if the gate would overlap over the contact region of length $L_C$ (Fig. 1). For a single-gate (SG) single-sheet device, this can simply be enabled by having the gate contact on the side opposite to the contacts, i.e., using for instance a top-contact and an individually back-gated transistor [1]. To enable a D2 tri-gate with stacked sheets, however, we propose here a doubled forked structure (E2), where the sheets are connected to a forked gate on one side and to forked S & D contacts on the other side (Fig. 1). Our simulation results show, as expected, that such a multigate E2D2 architecture enables a better electrostatic control and improved drive current at scaled CGP, especially for Si where the film thickness can be relaxed, compared to the SG-D2 transistor. We report here on the impact of the multigate E2D2 architecture innovation on intrinsic-device and loaded-invertor performance, when pitch is scaled well below 30 nm using accurate dissipative DFT-NEGF atomistic-simulation fundamentals and cell-layout extrinsics. The E2D2 architecture is benchmarked to NS MOSFETs using both Si and 2 emerging 2D transition metal dichalcogenide (TMD) monolayer (1ML) materials – one, $WS_2$, with predicted fundamental drive similar to that of Si, the other, $HfS_2$, featuring an enhanced fundamental drive current [1] – as test vehicles.

## Methods

Current – Voltage ($I_D(V_G)$, Fig. 2d inset) and intrinsic device capacitances ($C_{Gi}$) (Fig. 2b) for Si and 2D TMD E2D2 and standard NS references are simulated using our first-principle atomistic NEGF solver ATOMOS, including electron-phonon scattering [1,2]. From these simulations the intrinsic single-sheet device fundamental performance vs. CGP can be assessed (Fig. 2). For each CGP, a full device optimization is made including film thickness ($t_S$) scaling for Si and extension doping for the NS. Note that the detrimental impact of quantum confinement, including mobility degradation, and source-to-drain direct tunneling are naturally included in our quantum transport solver. For computing stacked-invertor energy-delay products (Fig. 3), the extracted extrinsic capacitance of the cell layout $C_{cell}$ and the backend-of-line load, $C_{BK}$ are used (Fig. 1). $C_{cell}$ values are reported in Fig. 4a. The number of stacked sheets ($n_S$) used is computed to allow a total stack height of 60 nm for all devices. $n_S$ is the same for E2D2 and NS of a same material ($n_S = 4$ for Si and 5 for the TMDs owing to their 1ML thickness of about 0.6 nm [1]). The available width for a single sheet, W, in our 5-track E2D2 layout cell is 12 nm. The standard NS layout is described in [3] and W is 12 nm as well.

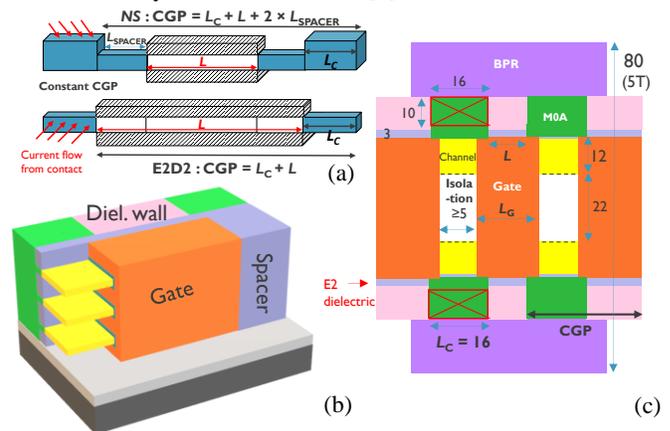

Fig. 1 E2D2 device structure. a) Side view schematic of a single-sheet multigate conventional NS (Top) and E2D2 transistor (Bottom) with same CGP. b) 3D view showing the doubled forked (E2) structure, c) cell layout of the 5-track (cell height = 80 nm) E2D2 invertor cell with buried power rail (BPR). The technological dimensions, we assumed for this study are indicated in the figure. We assumed $L_G = L$. $L_C = 16$ nm, $L_{SPACER} = 5$ nm. The width of an individual sheet is $W = 12$ nm. The P/N separation is 22nm. For the gate oxide, we assumed a 2 nm hafnium oxide with $\varepsilon_R = 15.6$. The gate stack metal thickness is 6 nm. The channel (white region in Fig. 1a) is intrinsic. The contact regions (in blue in Fig. 1a) are doped. $n_S = 4$ for Si and 5 for 2D.

## Results

Owing to its 10 nm extended gate length at same CGP, the E2D2 SS and, hence, $I_{ON}$ at fixed $I_{OFF}$ are superior compared to that of the NS as CGP is scaled below 30 nm for all materials. The E2D2 $C_{Gi}$ are however larger at fixed CGP, the net effect being that the **E2D2 optimal intrinsic delay is comparable to that of its NS counterpart, but shifted towards smaller**

**CGPs by about 10 nm, i.e., $2 \times L_{SPACER}$ (Fig. 2). Hence the E2D2 architecture enables a significant scaling boost.** For Si, the optimal NS and E2D2 delays are obtained at CGP = 36, and 26 nm, respectively, i.e., $L_G$ = 10 nm and $t_S$ = 3 nm in both cases. For the 2D materials, a further 5 nm scaling boost is observed, and optimal delays are achieved at CGP = 31 and 21 nm for the NS and E2D2 respectively, corresponding to $L_G$ = 5 nm in both cases. For the TMDs the 21 nm E2D2 CGP corresponds to the case where CGP is only limited by the contacts ($L_C$ and the minimum isolation spacing, IS, required to separate subsequent pads, assuming IS = $L_{SPACER}$ (Fig. 1)), hence ultimate gate scaling has been achieved. Further CGP reduction could only be achieved by scaling the contacts.

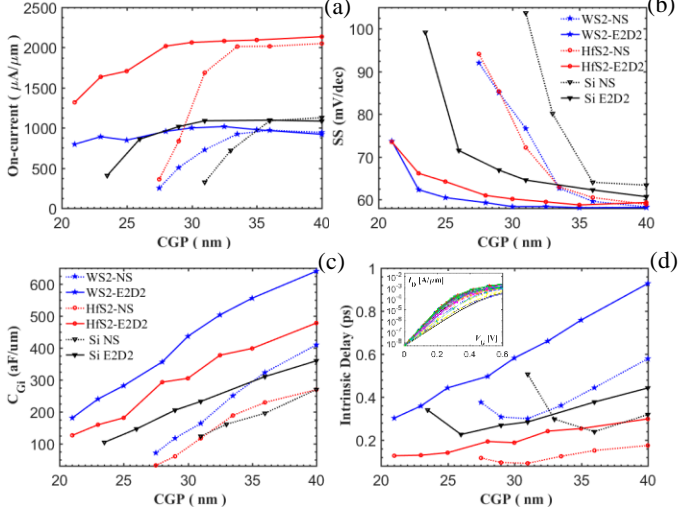

Fig. 2 a) On-current ($I_{ON}$) at fixed $I_{OFF}$, b) subthreshold slope (SS), c) intrinsic gate capacitance ($C_{Gi}$) and d) intrinsic delay - assuming an effective current $i_{eff}$ = 0.45 × $I_{ON}$ – vs. CGP for Si and for WS$_2$ and HfS$_2$ 2D monolayers NS and E2D2 architectures from ab-initio - NEGF transport simulations ( the trace of the $I_D(V_G)$ characteristics are in the inset) [1]. For the NS, $L_G$ = CGP - $L_C$ - $2 \times L_{SPACER}$, while for E2D2 $L_G$ = CGP - $L_C$. $L_C$ = 16 nm, $L_{SPACER}$ = 5 nm. $I_{ON}$ is normalized by the gate perimeter. $I_{OFF}$ = 5 nA/µm. $V_{DD}$ = 0.6V.

Next, we investigate the switching energy vs. delay (EDP) of high-performance stacked E2D2 and NS loaded inverters for different CGPs at various $V_{DD}$ (Fig. 3). For HfS$_2$ E2D2 and NS invertors, the optimal EDP is achieved at CGP = 21, $L_G$ = 5 nm and CGP = 31, $L_G$ = 5 nm respectively. We obtain a similar result for the WS$_2$ case (not shown here). For Si E2D2 and NS invertors, the optimal EDP is achieved at CGP = 26, $L_G$ = 10 nm and CGP = 36, $L_G$ = 10 nm respectively. Any further attempt to scale CGP by scaling $L_G$ beyond this optimal value results in significant performance reduction (for the TMD E2D2 it is simply not possible to further scale CGP with $L_G$ scaling). **These results further confirm the 10 nm improved scalability, we obtained from the intrinsic device performance and delays (Fig. 2). The E2D2-inverter improved EDP performance compared to that of the NS is mostly linked to the reduced $C_{BK}$ owing to CGP scaling.**

This is confirmed in Fig. 4, where the loaded-invertor EDPs are shown for the 3 material cases at their optimal CGP and $L_G$ values with and without $C_{BK}$ included in the load. Regardless of the material system used, without $C_{BK}$ (Fig. 4.a), the E2D2 performance are similar to that of their NS counterparts, while they are enhanced when $C_{BK}$ is included (Fig. 4.b) (the E2D2 and NS devices also share the same optimal $L_G$ of 10 nm for Si and 5 nm for the TMDs).

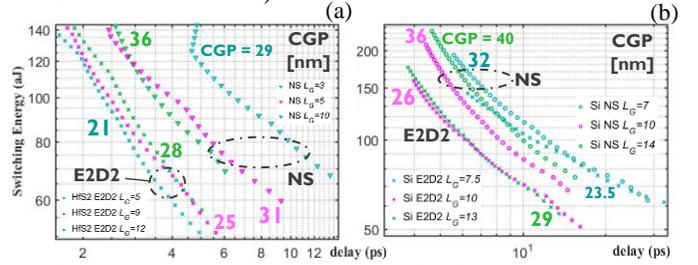

Fig. 3 Switching energy vs. delay (EDP) of high-performance stacked E2D2 and NS inverter cells for different CGP and $L_G$, as indicated in the figures, at various $V_{DD}$ (0.4V to 0.7V). The devices are made of a) 1ML-HfS$_2$ with $n_S$ = 5 sheets/device, b) Si with $n_S$ = 4 sheets/device and optimized Si thickness $t_S$ ranging from 3 to 5 nm. The inverters are loaded with the extrinsic capacitances of the cell layout $C_{Cell}$ and a 50 CGP-long metal line with capacitance $C_{BK}$ = 198 aF/µm [4]. $I_{OFF}$ = 5 nA/µm.

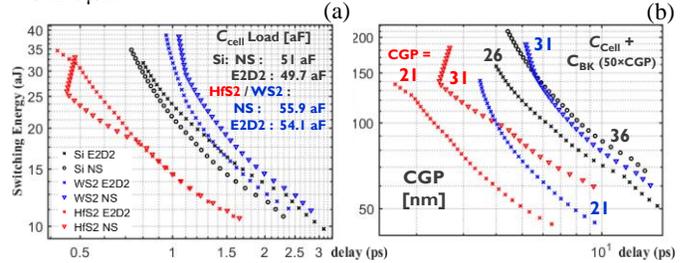

Fig. 4 EDP of the Si, WS$_2$ and HfS$_2$ stacked E2D2 and NS inverter cells at optimal CGP as indicated in Fig. 4b, at various $V_{DD}$ (0.4V to 0.7V). The inverters are loaded with a) $C_{cell}$ only, its value for each invertor case is indicated in the figure, b) $C_{cell}$ and a 50 CGP-long metal line $C_{BK}$. $I_{OFF}$ = 5 nA/µm.

## Conclusions

We proposed a compact E2D2 multigate architecture that enables sub-30-nm CGP, i.e., an improved $2 \times L_{SPACER}$ pitch scaling, compared to a NS reference, owing to the suppression of ungated extensions from the CGP equation. This E2D2 scaling benefits were measured in term of similar intrinsic performance and optimal delay but at a 10 nm reduced CGP. A similar conclusion was found comparing E2D2 and NS stacked-invertor cells. For backend-loaded invertors, the E2D2 EDP performance is further enhanced due to CGP and, hence, $C_{BK}$ reduction. Similar relative benefits were observed regardless of the material system used. Compared to Si, a mature 2D material technology could potentially further enable an extra 5-nm CGP scaling boost, both for the E2D2 and NS architectures, with same or improved performance, if respectively WS$_2$, a material with a fundamental drive similar to Si, or HfS$_2$, a higher mobility material, were used.

### References
[1] A. Afzalian, *npj 2D Mater Appl* **5**, 5 2021. [2] A. Afzalian, *IEEE Trans Electron Devices*, 68,11,2021. [3] Z. Ahmed, *IEDM, 2020*, pp. 22.5.1. [4] https://irds.ieee.org/editions/2018